# Scientifically unacceptable concept of the Epoch of Malthusian Stagnation

Ron W Nielsen aka Jan Nurzynski[1]

Environmental Futures Centre, Gold Coast Campus, Griffith University, Qld, 4222, Australia

October, 2013

In order to control the growth of human population it is helpful to understand correctly the mechanism of growth, and the first essential step is to investigate current interpretations and reject any unscientific explanations. One of such popular but questionable interpretations is the concept of the Epoch of Malthusian Stagnation. We discuss its origin, narrative and claims. We explain why this concept is scientifically unacceptable. This investigation questions also the closely-related Demographic Transition Theory, whose essential component is the assumed mechanism of Malthusian stagnation for the first stage of growth.

**Introduction**[2]

Seven groups of trends are now shaping the future of our planet (Nielsen, 2005, 2006, 2007). One of them, and the prime mover of the remaining six, is the excessive growth of human population combined with our unique, insatiable drive to consume more than we need to support our life.

---

[1] r.nielsen@griffith.edu.au; ronwnielsen@gmail.com; http://home.iprimus.com.au/nielsens/ronnielsen.html

[2] Discussion presented in this publication is a part of a broader study, which is going to be described in a book (under preparation): *Population growth and economic progress explained*.

The growth of human population is likely to reach a maximum and start to decline, but even if it does we cannot be sure that it will not start increasing again to repeat the currently experienced runaway process. In fact, close inspection of data (Manning, 2008; US Census Bureau, 2013) strongly suggests that such a repeated runaway process is not only possible but also probable.

If we survive, it will be essential for us to learn how to control the growth of human population, and we can have a better chance of doing it if we can understand correctly the mechanism of growth, if we can identify and understand its driving forces, if we can explain why the population was growing so slowly in the distant past and why its pace increased so significantly in the last 200 years. It is in our interest to be on guard against accepting incorrect ideas and misconceptions about the driving forces of growth because such ideas divert our attention from the correct interpretation of the mechanism of growth and from finding a successful solution to the key problem of controlling the growth of population.

One of such misconceptions is the postulate of Malthusian stagnation, the mechanism, which is supposed to have controlled the growth of population over thousands of years and creating the Epoch of Malthusian Stagnation also known as the Malthusian Regime, the concept, which gained surprisingly high popularity even though a much simpler and elegant explanation is suggested by the data (Maddison, 2010; Manning, 2008; US Census Bureau, 2013) and by the yet insufficiently explored evidence that they can be described using hyperbolic distributions (Shklovskii, 1962, 2002; von Foerster & Amiot, 1960; von Hoerner, 1975).

The concept of Malthusian Stagnation gradually evolved into a system of ideas and explanations that could be described as a myth because it is based on beliefs and fragments of questionable information, all fused together and shaped by a good dose of creative imagination, by filling in the gaps and by adding new twists to the already interesting story, the story that has been told and retold many times without seriously testing its validity, maybe because like many other good stories, this narrative is also too good and too attractive to be spoiled by facts.

This concept is reinforced by subsidiary ideas, postulates, mechanisms and explanations invented on the run and expected to be accepted by faith, claims unsupported by a rigorous examination of empirical evidence and yet proclaimed with an absolute certainty. Whatever is convenient to make the story attractive is promptly invented and stated with confidence. We shall discuss a few examples of such ideas and explanations but more can be found easily and readily in the relevant and unfortunately abundant literature.

Two important elements make the concept of the Epoch of Malthusian Stagnation deceptively attractive: it is strongly believable and it is over 200 years old. It is believable because the growth of human population over thousands of years was indeed slow, so slow that it appears to have been stagnant. It is also an old concept because its origin can be traced to Malthus (1798).

> The Malthusian theory, as was outlined initially by Malthus (1978), captures the main attributes of the epoch of Malthusian stagnation that had characterized most of human existence… (Galor, 2005, p. 221)

> The idea of multiple equilibria, or poverty traps, can be retraced back to Malthus (Wang, 2005, p. 36).

The work of Malthus was the first documented attempt to understand and explain human population dynamics. Considering the time it was written, it was a notable achievement, but it should have been developed further into a better-structured scientific paradigm.

**The first tentative steps**

Malthus worked under difficult conditions because he did not have access to the vast amount of information available to us. His pioneering work is important because he has pointed out to the limits of growth and warned against the danger of the excessive growth of human population.

He imagined that food production is linear but that the growth of human population exponential, at least in principle, because according to him it is also controlled and shaped by positive and preventive checks. Malthus does not write much about the preventive checks, which he imagines only as a prudent postponement of marriage, but he elaborates a little more about the positive checks.

> Famine seems to be the last, the most dreadful resource of nature. The power of population is so superior to the power in the earth to produce subsistence for man, that premature death must in some shape or other visit the human race. The vices of mankind are active and able ministers of depopulation. They are the precursors in the great army of destruction; and often finish the dreadful work themselves. But should they fail in this war of extermination, *sickly seasons, epidemics, pestilence, and plague, advance in terrific array, and sweep off their thousands and ten thousands*. Should success be still incomplete, *gigantic inevitable famine stalks in the rear, and with one mighty blow levels the population* with the food of the world.

> Must it not then be acknowledged by an attentive examiner of the histories of mankind, that in every age and in every state in which man has existed, or does now exist.
>
> That the increase of population is necessarily limited by the means of subsistence.
>
> That population does invariably increase when the means of subsistence increase. And that the superior power of population is repressed, and the actual population kept equal to the means of subsistence, *by misery and vice*? (Malthus, 1798, p. 44. Italics added.).

The last sentence in this quotation indicates that Malthus understood positive checks in a broader sense. His concept includes more than the generally mentioned cataclysmic events such as famines, pestilence and wars.

> Through the animal and vegetable kingdoms, nature has scattered the seeds of life abroad with the most profuse and liberal hand. She has been comparatively sparing in the room and the nourishment necessary to rear them. The germs of existence contained in this spot of earth, with ample food, and ample room to expand in, would fill millions of worlds in the course of a few thousand years. Necessity, that imperious all pervading law of nature, restrains them within the prescribed bounds. The race of plants and the race of animals shrink under this great *restrictive law*. And the race of man cannot, by any efforts of reason, escape from it. *Among plants and animals its effects are waste of seed, sickness, and premature death.* Among mankind, *misery and vice*. The former, misery, is an absolutely necessary consequence of it. Vice is a highly probable consequence, and we therefore see it abundantly prevail, but it ought not, perhaps, to be called an absolutely necessary consequence. The ordeal of virtue is to resist all temptation to evil (Malthus, 1798, p. 5. Italics added.).

So it appears that by misery Malthus understood all forms of deprivation and by vice the way humans respond to misery. He writes, for instance, "the actual distresses of some of the lower classes, by which they are disabled from giving the proper food and attention to their children, act as a *positive check* to the natural increase of population" (Malthus, 1978, p. 20. Italics added.). He also adds: "the distresses which they suffer from the want of proper and sufficient food, from hard labour and unwholesome habitations, must operate as a *constant check* to incipient population" (Malthus, 1978, p. 31. Italics added.). Positive checks include, therefore, not only such large demographic catastrophes as famines, pestilence and wars but also various forms of adverse living conditions. In order to understand the relation between positive checks and the growth of human population we can study not only the effects of demographic catastrophes in the past but also the effects of adverse living conditions experienced now in many countries.

Malthus paved the way towards a correct understanding of population dynamics but all his claims and explanations should not have been ever accepted without testing them, and we now have many ways of doing it. If confirmed and supported by closer examination, we can use them. This is the accepted process of scientific investigation. We can check whether the population was increasing exponentially. We can check whether food production was increasing linearly. We can study the impacts of demographic catastrophes and see whether they were shaping the growth of human population. We can study the impacts of harsh living conditions on the growth of human population, impacts such as hunger, poverty and infectious diseases. What Malthus could not have done because he did not have access to the vast empirical evidence available to us, we certainly can do it now. He was opened to suggestions and corrections. Referring to himself in the third person he wrote:

> If he should succeed in drawing the attention of more able men to what he conceives to be the principal difficulty in the way to the improvement of society and should, in consequence, see this difficulty removed, even in theory, *he will gladly retract his present opinions and rejoice in a conviction of his error* (Malthus, 1798, p. viii. Italics added.)

Scientific research calls for a high level of integrity. Malthus could have only said that he would "gladly retract his present opinions" but he went a step further and added that he would "rejoice in a conviction of his error," the declaration revealing his unbiased attitude to his own concepts, an example every scientist should follow.

Malthus is known for suggesting positive and preventive checks but he is not so well known for his concept of an *efficient replacement mechanism.*

> The absolute population at any one period, in proportion to the extent of territory, could never be great, on account of the unproductive nature of some of the regions occupied; but there appears to have been a most rapid succession of human beings, and *as fast as some were mowed down by the scythe of war or of famine, others rose in increased numbers to supply their place*. Among these bold and improvident Barbarians, population was probably but little checked, as in modern states, from a fear of future difficulties (Malthus, 1798, p. 15. Italics added.).

So, if we read closely what Malthus wrote we can see that he did not propose just one or two mechanisms of growth but three.

1. Malthusian stagnation mechanism: The *spontaneous*, *growth-suppressing* mechanism induced by positive checks.

2. Malthusian preventive checks mechanism: The *voluntary*, *growth-suppressing* mechanism controlled by preventive checks.
3. Malthusian replacement mechanism: The *spontaneous*, *growth-promoting* mechanism, i.e. the *replacement mechanism,* acting in the presence of positive checks.

The Malthusian stagnation mechanism and Malthusian replacement mechanism describe diametrically different effects of positive checks, the growth-suppressing and the growth-stimulating effects. One of these mechanisms must be inapplicable unless we accept the implied explanation proposed by Malthus that the second mechanism applies only to improvident "Barbarians, brave, robust, and enterprising, inured to hardship, and delighting in war" (Malthus, 1978, p.15) while the first mechanism applies to "higher classes" and to people whose manners are "pure and simple" (Malthus, 1978, p. 9).

An alternative and more convincing explanation is that there is no such thing as the Malthusian stagnation mechanism. The growth of population is controlled by some kind of a steady mechanism, which can be slowed down by purposefully applied preventive checks. However, stagnation is impossible because positive checks trigger automatically the replacement mechanism, which quickly repairs the damage caused by positive checks.

It would be hard to prove that the convoluted implied explanation of the two mechanisms of growth for two different groups of people is correct but it would be much easier to disprove it by studying the growth of population in poor countries where the positive checks are still active to see whether the replacement mechanism is also active and whether it is limited only to improvident Barbarians delighting in war. Indeed, rather than focusing entirely on the Malthusian stagnation mechanism, which probably does not work, we should also study the Malthusian replacement mechanism. We should combine this study with a close examination of data (Maddison, 2010; Manning, 2008; US Census Bureau, 2013). We should try to explain why the apparent natural tendency of the growth of population is to follow hyperbolic trajectory (Shklovskii, 1962, 2002; von Foerster & Amiot, 1960; von Hoerner, 1975), the amazingly simple pathway, suggesting that the assumption of the Malthusian stagnation mechanism is irrelevant and inapplicable

It is interesting that Malthus used arithmetic and geometric progressions to support his arguments but it is not certain whether he was familiar with the hyperbolic growth, let alone that he appreciated the difference between the hyperbolic and exponential (geometrical) types of growth. Even now many people do not know the difference, and as discussed in another place (Nielsen, 2013a, 2013b, 2013c) they fall into the trap of the hyperbolic illusion, the deception persuading them to accept the concept of the Epoch of Malthusian Stagnation and the associated concept of the escape from the mythical Malthusian trap or a similar concept of a sudden intensification in the growth of human population (Johnson & Brook, 2011).

**The stagnation**

One would expect that in the course of time, the explanations and ideas put forward by Malthus would have been checked by data and if necessarily corrected, but in over 200 years, little progress, if any, has been made in this field, and whatever progress has been made, such as in showing that the growth of human population is not exponential but hyperbolic and remarkably stable (Shklovskii, 1962, 2002; von Foerster & Amiot, 1960; von Hoerner, 1975), has been generally ignored. The original ideas of Malthus about the effects of positive checks have been adorned by many colourful and attractive descriptions until they evolved into a powerful and compelling narrative. *If there is any form of stagnation, it is the stagnation in the understanding of the human population dynamics.*

> The history of population theory can be summarized in three words: pre-Malthusian, Malthusian, and post-Malthusian. Hardly ever in intellectual history does one man so dominate a field as does the Reverend Thomas Robert Malthus in demographic theory. To paraphrase a quotation attributed to Newton, Malthus' shoulders *must* be climbed (Thomlinson, 1965, p. 47. Italics in the original text.).

…the demographic transition experiences three regimes: the 'Malthusian Regime,' the 'Post-Malthusian Regime,' and the 'Modern Growth Regime.' *Any theory* attempts (sic) to describe the process of demographic transition *must* include these three periods (Wang, 2005, p. 3. Italics added.).

Claiming, suggesting or assuming that something *must* be accepted just because it comes from a certain source is not acceptable in science. Any theory can be questioned and even should be questioned, and if necessarily corrected or rejected. The sooner it is done the better. If Malthus's shoulders must be climbed it is only for the same reason as climbing the shoulders of any giants of human intellect: to see better and further ahead.

**The myth**

According to the concept of the Epoch of Malthusian Stagnation, human population was locked in the Malthusian Trap of positive checks for many thousands of years, the trap controlling and suppressing growth, the process reflected in fluctuations or random oscillations in the size of human population (Galor 2005, 2007; Galor & Moav, 2001; Galor & Weil 1999, 2000; Manfredi & Fanti, 2003). The growth was slow, if any, chaotic and unpredictable. This narrative reflects closely the descriptions of the first stage of growth claimed by the Demographic Transition Theory (Caldwell, 1976, 2006; Casterline, 2003; Coale, 1973; Haupt & Kane, 2005; Kirk, 1996; Landry, 1934; Lee, 2003; Lehr, 2009; Notestein, 1945; Olshansky & Ault, 1986; Olshansky, Carnes, Rogers, & Smith, 1997, 1998; Omran, 1971, 1983, 1998, 2005; Rogers & Hackenberg, 1987; Singha & Zacharia, 1984; Thompson, 1929; van de Kaa, 2008; Warf, 2010).

While the origin of the concept of stagnation can be traced back to Malthus and linked with the well-known Demographic Transition Theory, the term "epoch of Malthusian stagnation" was probably first introduced by Galor and Moav (2001) and strongly reinforced by Galor in his so-called Unified Growth Theory (Galor, 2004, 2005), the theory producing no verifiable distributions and not a single fit to the data. Indeed, in the detailed discussion of this theory (Galor, 2005), the discussion containing over one hundred pages of closely-spaced print, many complicated formulae, calculations and graphs are presented but not a single graph comparing theoretical predictions with the relevant data (Maddison, 2001), the data referred to in this document but not used for a direct verification of the discussed theory.

A broader concept attempting to explain the growth of human population is the idea of the existence of *three* regimes of growth: (1) Malthusian Regime (or Malthusian Epoch), (2) Post-Malthusian Regime and (3) Sustained (or Modern) Growth Regime (Galor & Weil, 1999, 2000; Galor, 2005), the idea *contradicted* by the close analysis of data (Maddison, 2001) known to Galor (2005). This proposed sequence of growth and the sequence claimed by the Demographic Transition Theory, containing *four* stages of growth but maybe even *five* (Haupt & Kane, 2005; Olshansky, Carnes, Rogers, & Smith, 1998; Schmid, 1984; van de Kaa, 2008), or *six* stages (Myrskyla, Kohler & Billari, 2009), is too untidy and unappealing. Each stage of growth is governed by different sets of forces. (Routinely more than one force is assumed for each stage.) In addition, certain specific forces have to be assumed for each transition between relevant stages, all this creating a complex explanation of growth, while the data (Maddison, 2010; Manning, 2008; US Census Bureau, 2013) and their limited analysis (Shklovskii, 1962, 2002; von Foerster & Amiot, 1960; von Hoerner, 1975) suggest a *simpler mechanism* and a better explanation.

Rather than moving forward we seem to be moving in circles. Rather than looking for an alternative and possibly simpler and more suitable interpretation of the growth of human population suggested both by the data and by their limited analysis, the interpretation unknown to Malthus, because his work was based on strongly limited information, we seem to be trapped by focusing strongly on just one mechanism of growth he has considered, the mechanism of stagnation that probably never worked. What is probably simple is made complicated and untidy.

The problem with the explanation of the human population dynamics reminds about the problem encountered many years ago with the explanation of the dynamics of celestial bodies. Describing the work of mathematicians of his time, Osiander wrote:

> With them it is as though an artist were to gather the hands, feet, head and other members from his images from divers models, each part excellently drawn, but not related to a single body, and since they in no way match each other, the result would be monster rather than man (Copernicus, 1995).[3]

During the long-lasting, mythical Epoch of Malthusian Stagnation, birth rates are claimed to have been high because new generations were needed to support many tiresome and mundane activities such as hunting, gathering, cultivating crops, caring for children and generally for coping with harsh living conditions.

> According to Classical economists, and early Neo-Classical economists as well, population size was determined by the demand for labor. This was the Law of

---

[3] This quotation comes from a letter written by Andreas Osiander, Lutheran theologian and a friend of Copernicus, a letter addressed to the chief editor, Pope Paul III. Osiander argues in favour of the mathematically simple and elegant heliocentric system as opposed to the complicated geocentric descriptions. This letter was later used as an unsigned introduction to the book *De revolutionibus orbium coelestium*, and was mistakenly attributed to Copernicus.

Population which constantly operated behind the seemingly random variations in fertility and mortality induced by epidemic, famine, and war (Lee, 1997, p. 1063).

Claims:

1. Population size was determined by the demand for labour
2. This is the Law of Growth
3. This law has been accepted by Classical and early Neo-Classical economists
4. There were seemingly random variations in fertility and mortality
5. Random variations were caused by epidemics, famine and war
6. This law operated constantly behind these seemingly random variations.

It is interesting how much is claimed in this single paragraph and it does not matter whether Lee agrees with all these claims or just describes them. This quotation represents a typical set of questionable claims often encountered in publications related to the concept of the Epoch of Malthusian Stagnation. Can we prove them or do we have to take them by faith?

To prove this "Law of Population" we would have to have data about the demand for labour and about the growth of population extending over thousands of years, and we would have to prove that there is a correlation between the demand for labour and the size of human population, or the birth rates. We cannot prove it because we do not have such data, but we can show that the population data (Maddison, 2010; Manning, 2008; US Census Bureau, 2013) do not display any features that could linked with this "Law of Population." Was human procreation really guided so rationally by the demand for labour or was it prompted by more basic and primordial force? There is nothing in the population data to support this "Law of Population" and nothing to support the claims of "Classical economist, and early Neo-Classical economists as well."

It is easy to accept, without a proof, that there were random variations in the fertility and mortality. It would be probably more difficult to expect that there were no variations but we have no information about the amplitude of these variations because while we have reliable data about the *size* of human population (Maddison, 2010; Manning, 2008; US Census Bureau, 2013) over thousands of years we have *no matching data about fertility and mortality (birth rates and death rates).*

How can we ever claim that these assumed and imagined random variations were "induced by epidemic, famine, and war"? How can we feel safe in taking such a leap of faith? How can we expect that such leaps of faith will lead us in the right direction? The only outcome we can expect is that they will lead us gradually further away from finding correct answers.

It should be also noted that the growth of population is not determined by the *absolute* values of birth and death rates but by the *difference* between these two quantities. This difference determines the *growth rate*. A constant difference (growth rate) produces *exponential* growth. A zero difference produces *constant* population. However, variable difference (growth rate) does not necessarily produce a variable *size* of the population and we shall investigate this issue further in the next publication. Even if the birth and death rates were high and fluctuating we cannot automatically claim that they were producing random fluctuations and stagnation in the size of human population.

According to the concept of the Epoch of Malthusian Stagnation, as soon as the population started to increase, it was slowed down or significantly reduced by numerous factors associated with harsh living conditions.

> During the first [stage of the demographic transition], fertility is assumed to have been sufficiently high to allow a population to grow slowly even in the face of a rather high level of mortality. However, periodic epidemics of plague, cholera, typhoid and other infectious diseases would *in one or two years wipe out the gains made over decades. Over long periods of time there would, consequently, be almost no population growth at all* (van de Kaa, 2008).

Claims:

1. During the first stage of the demographic transition fertility and mortality are assumed to have been high
2. Population was growing slowly
3. Population growth was strongly controlled by periodic epidemics of plague, cholera, typhoid and other infectious diseases
4. Periodic epidemics of plague, cholera, typhoid and other infectious diseases would in *one or two years* wipe out the gains made over *decades*
5. Over long periods of time there was no population growth at all

Van de Kaa describes the first of four stages of growth claimed by the classical Demographic Transition Theory, the stage corresponding to the mythical Epoch of Malthusian Stagnation. Here we have a vivid description of what was happening so long ago and over a long time, not only a vivid description but also an explanation, as if we moved back in time and saw it all happening in front of our eyes. However, this account is in direct contradiction with the data describing the growth of human population (Maddison, 2010; Manning, 2008; US Census Bureau, 2013) because there were never "long periods of time" when there was "almost no population growth at all." It can be easily checked using the data that the growth of population, global, regional and even local, was in general following remarkably stable trajectories.

It is both amazing and disturbing that these hypothetical chaotic changes in the growth of human population are not only so confidently claimed but also so categorically explained by correlating them with "periodic epidemics of plague, cholera, typhoid and other infectious diseases." It would be hard, or impossible, to demonstrate these correlations: hard because one would have to analyse records of all demographic catastrophes and try to isolate the impacts caused by "periodic epidemics of plague, cholera, typhoid and other infectious diseases;" hard or impossible because it is generally hard or impossible to isolate specific causes of death; impossible because population data (Maddison, 2010; Manning, 2008; US Census Bureau, 2013) do not show any signs of periodic crashes and recoveries in the growth of population. It is impossible to correlated the non-existent features with "periodic epidemics of plague, cholera, typhoid and other infectious diseases." Typically for such confident claims, if they are not closely scrutinised they might sound attractive and convincing, but they have to be accepted by faith.

We seem to know also so much about the birth and deaths rates, how high they were and how they were fluctuating for thousands of years but all these descriptions, pronounced with confidence, are based on speculations and conjuncture because *we simply do not have the relevant data* to support these claims. We may consider ourselves fortunate to have fairly reliable estimates of the *size* of human population in the distant past (Maddison, 2010; Manning, 2008; US Census Bureau, 2013) but we have *no matching data for the birth and death rate*s. The data for the size of human population do not support the concept of stagnation.

We might feel or think that our descriptions are true; we might wish that they were true, but we should test them by empirical evidence. Furthermore, even if we assume that birth and death rates were high and strongly fluctuating we cannot automatically claim that such fluctuations are reflected in the *size* of human population. We might feel that they are but we would have to prove it. All these speculations about the death and birth rates being high, closely balanced and producing stagnant state of growth are not based on solid scientific evidence and on the accepted process of scientific investigation but on leaps of faith reinforced by creative imagination.

We also seem to have so much information about the harsh living conditions in the distant past and about their suppressive influence on the growth of human population but we are ignoring the contradictory evidence in the third-world countries. If we spent more time on investigating empirical evidence rather than on creative writing maybe we could learn something useful about human population dynamics.

> …the food-controlled homeostatic equilibrium had prevailed since time immemorial" (Komlos, 2000, p. 320).

> …the population tends to oscillate in a homeostatic mechanism resulting from the conflict between the population's natural tendency to increase and the limitations imposed by the availability of food (Artzrouni & Komlos, 1985, p. 24).

Claims:

1. There was a food-controlled homeostatic equilibrium
2. This equilibrium prevailed since time immemorial
3. Population tends to oscillate in a homeostatic mechanism
4. Oscillations are caused by the natural tendency of the population to increase and by the limitations imposed by the availability of food

There is nothing to stop anyone from assuming homeostatic mechanism for the growth of human population but to claim that "homeostatic equilibrium had prevailed since time immemorial" we would have to work a little harder. We would have to design a model with the homeostatic equilibrium and show *that it fits the relevant data* "since time immemorial," but even then we would have to allow for the possibility that some other mechanism could also fit the data equally well or maybe even better. Life would be too easy if we could just imagine that something happened and claim that it did happen. We have no convincing evidence that there was homeostatic equilibrium between the supply of food and the size of human population let alone that "food-controlled homeostatic equilibrium had prevailed since time immemorial."

Artzrouni and Komlos (1985) claim that "the population tends to oscillate in a homeostatic mechanism." Such oscillations add an extra degree of difficulty in reconciling the theory with the data. The oscillations should be produced by the model but even more importantly they should be also *demonstrated in the relevant data*.

Population data (Maddison, 2010; Manning, 2008; US Census Bureau, 2013) show no signs of such oscillations. Furthermore, if we examine closely the results of the calculations based on this "homeostatic mechanism" we shall see that Artzrouni and Komlos (1985) generated a steadily-increasing *exponential* growth with no signs of any oscillations and that their calculated distribution does not fit the population data.

The absence of the desired oscillations or stagnation and the disagreement with the data show that the assumed mechanism of Malthusian stagnation does not work. The claim that "the population tends to oscillate in a homeostatic mechanism" is neither confirmed by the data nor by the model, which assumes the presence of such oscillations.

> Stage 1 [of the Demographic Transition Theory] presumably characterizing *most of human history*, involves high and relatively equal birth and death rates and little resulting population growth" (Guest & Almgren, 2001; p. 621. Italics added.).

This stage is characterized not by changes in *average* death rates but by a *stagnation of death rates at extremely high levels* for a period of what is believed to be *thousands of years*" (Olshansky & Ault, 1986, p. 357. Italics added.).

Claims:

1. Stage 1 of the demographic transition presumably characterised most of human history
2. During this stage 1 there were high and relatively equal birth and death rates
3. During this stage there was little resulting population growth
4. This stage was not characterised by changes in the average death rates
5. This stage was characterised by stagnation of death rates at extremely high levels
6. This stagnation is believed to have lasted for thousands of years

It is amazing how many details we know or believe to know and how well we understand what was happening over thousands of years without having strong empirical evidence to support all these claimed details.

Birth and death rates may have been high and strongly fluctuating but it does not matter. High and fluctuating birth and death rates do not necessarily prove the existence of a stagnant state of growth because, as mentioned earlier, growth is determined by the average *difference* between these two quantities. (We shall examine this issue more closely in the next publication.) Even more importantly, studying just the death rates *or* birth rates, or equivalently studying just the fertility rates (Lehr, 2009) cannot be used as the evidence of stagnation or of the demographic transitions because if for instant the average fertility rates decrease in the same way as the average mortality rates, if the gap between them is approximately constant or gradually increasing, they will not produce any form of transition in the growth of population or any form of stagnation but rather a steady and undisturbed growth.

To generate a stagnant state we would have to have fertility and mortality rates changing in a very special way. The average difference between them could not be constant or increasing but it should be *zero*. While we cannot investigate the long-range time-dependence of the birth and death rates because we do not have the relevant data, we can study the time-dependence of the *size* of the population and these data do not confirm the existence of any form of stagnation, let alone stagnation that lasted for thousands of years.

> It is well documented that the fluctuations experienced by the world's population throughout history did not have a regular, cyclical pattern, but were, to a large extent, brought about by randomly determined demographic crises (wars, famines, epidemics, etc.). As McKeown and others have pointed out, the main cause of these fluctuations of the past were mortality crises. There are four kinds of crises: subsistence crises, epidemic crises, combined crises (subsistence/epidemic), and finally crises from other causes, which are mainly exogenous (wars, natural or other catastrophes)
>
> Crises followed by *periods of population decline* during which the nutritional status of the population improved gave rise to fluctuations which testify to the continued existence of the 'Malthusian trap': population would not grow beyond its carrying capacity for long, and when it did, the resulting overshoot was followed by a 'crash' (i.e. the positive checks such as diseases, famines, wars, etc.) (Artzrouni & Komlos 1985, p. 24. Italics added.).
>
> Claims:
>
>   1. There were fluctuations in the world's population throughout history

2. These fluctuations are well documented

3. It is well documented that these fluctuations did not have a cyclic pattern

4. It is well documented that these fluctuations were, to a large extent, brought about by randomly determined demographic crises (wars, famines, epidemics, etc.)

5. The main cause of these fluctuations were mortality crises

6. There are four types of crises

7. Crises were followed by periods of population decline

8. Population decline improved nutritional status

9. Fluctuations testify to the continuing existence of Malthusian trap

10. Population was repeatedly reaching its carrying capacity

11. Population would not grow beyond its carrying capacity for long

12. Population growing beyond its carrying capacity was reflected in overshoots

13. Overshoots were followed by crashes.

If it is so well-documented it would be interesting to see at least a few references to this important and fundamental research work, to see the *data* for these fluctuations "throughout history," to see a positive proof that the "the fluctuations experienced by the world's population throughout history" are *correlated* with "demographic crises (wars, famines, epidemics, etc.)," that they were "brought about by randomly determined demographic crises." It would be also interesting to see convincing evidence that population was reaching its carrying capacity, that "population would not grow beyond its carrying capacity for long," the convincing evidence of overshoots and crashes, evidence that crashes were associated with "positive checks such as diseases, famines, wars, etc.," the compelling evidence of the existence of Malthusian trap, the demonstration of "periods of population decline," the compelling proof that periods of population decline caused by demographic crises were improving nutritional status.

It is well documented (Maddison, 2010; Manning, 2008; US Census Bureau, 2013) that the growth of human population does *not* show fluctuations or random behaviour. It is well documented that the data show no signs of frequent overshoots and crashes, no signs of growth reaching its carrying capacity, no signs of the "continued existence of the 'Malthusian trap'," no evidence that the "population would not grow beyond its carrying capacity for long," and no "periods of population decline." All these colourful and dramatic descriptions associated with the narrative of the mythical Epoch of Malthusian Stagnation are not confirmed by the population data.

It is obvious, that demographic crises were often causing decline in the size of *local* populations, depending on their scale and depending on what we understand by a local crisis. Sometimes it might have been just a large death toll in a city, a part of a country, as for instance in China (Mallory, 1926), or maybe in the whole country or even extending over a few countries. However, a large death toll does not necessarily mean a significant impact on the growth of human population. A large death toll should not be immediately interpreted as a population decline; it could have been just a slower growth over a certain time. All these issues should be closely investigated by examining records of demographic catastrophes. To arrive at any reasonably supported conclusion we would have to do some work. However, we have no data showing that these local demographic crises were repeatedly causing fluctuations in the growth of regional or global populations. In fact, the data show remarkably stable growth of human population, unaffected by demographic crises.

The opening statement in the above quotation contains two interesting and characteristic elements, the elements occurring repeatedly in the descriptions of the concept of the Epoch of Malthusian Stagnation: (1) it makes a highly-questionable but confident declaration about the *existence* of certain features (in this case about the existence of fluctuations) and (2) it equally confidently *explains* them while ignoring empirical evidence. The normal progression is *first to observe* certain features and then try to *explain* them. We can also reverse the process: we can first *predict* the existence of certain features. However, to accept the prediction and the associated explanation, we would have to *demonstrate the existence* of the predicted features.

So in this case, we would have to show first that there were significant fluctuations in the birth and death rates or in the size of human population and then we would also have to explain them convincingly by demonstrating that they were correlated with demographic crises. Alternatively, we would have to predict (using a suitable mathematical model) fluctuations in birth and death rates or in the size of human population by assuming that they are correlated with demographic crises and then we would have to show that our prediction is confirmed by the relevant data.

We cannot prove that there were fluctuations "throughout history" in the birth and death rates because we do not have the relevant data, but we can prove that there were no fluctuations "throughout history" in the *size* of human population because we have the relevant data (Maddison, 2010; Manning, 2008; US Census Bureau, 2013). There is nothing here to explain, except perhaps to explain the *absence* of fluctuations, the absence of random behaviour, crashes, overshoots or "periods of population decline."

Referring to three sources (Habakkuk, 1953; Kunitz, 1983; McKeown, 1983), Komlos explains:

> Malthusian positive checks (mortality crises) maintained *a long-run equilibrium between population size and the food supply*. Crises followed by periods when human nutritional status was above the level of subsistence gave rise to *cycles. …the cycles* *testify to the continued existence of the 'Malthusian population trap': population could not grow beyond an upper bound imposed by the resource and capital constraints* of the economic structure in which it was imbedded. The *'escape' from this trap* occurred only when the aggregate capital stock was large enough and grew fast enough to provide additional sustenance for the population, which thereby overcame the effects of the diminishing returns that had hindered human progress during *the previous millennia*. After escaping from the Malthusian trap, population was able to grow unchecked. In historic terms, this escape corresponds to the industrial and demographic revolutions. Removal of the nutritional constraint, at least for the developed part of the world, resulted in the population explosion (Komlos, 1989, pp. 194, 195. Italics added.).

Claims:

1. There was a long-term equilibrium between population size and the food supply
2. This equilibrium was maintained by positive checks (mortality crises)
3. Crises were followed by periods when human nutritional status was above the level of subsistence
4. This process gave rise to cycles
5. The cycles testify to the continued existence of the 'Malthusian population trap'
6. Population could not grow beyond an upper bound imposed by the resource and capital constraints of the economic structure in which it was imbedded
7. Malthusian trap was active for millennia
8. The escape for the Malthusian trap occurred when the aggregate capital stock was large enough and grew fast enough to provide additional sustenance for the population

Massive amount of work would be required to support all these impressive declarations. We would have to study food supply over millennia and determine how they were correlated with the growth of human population. We would have to prove that there was "a long-run equilibrium between population size and the food supply." We would have to study mortality crises over millennia. We would have to establish a correlation between the growth of human population, food supply and mortality crises. We would also have to investigate upper bounds of "resource and capital constraints" and prove that over millennia the size of the population was repeatedly reaching the limits of these upper bounds. It is easy to declare so much so quickly and with such a confidence, but it is harder to prove it. It is also hard to accept it, but accept we must if we want to accept the concept of the Epoch of Malthusian Stagnation.

The cycles cannot possibly testify to "the continued existence of the 'Malthusian population trap'" because they did not exist. Judging from the context, the reference here is to the cycles in the size of human population. There are no signs of cycles in the population data (Maddison, 2010; Manning, 2008; US Census Bureau, 2013). The absence of cycles and the steady growth of human population testify that the Malthusian trap did *not* exist. We cannot also claim that there was "'escape' from this trap" because there was no trap.

If the cycles refer to the mortality rates, it is even worse because we do not have the relevant data extending over millennia to claim that they "testify to the continued existence of the 'Malthusian population trap'."

Discussing the first stage of the Demographic Transition Theory, Warf explains:

> Because both fertility and mortality rates are high, the *difference* between them — natural population growth — is relatively low, *fluctuating around zero*" (Warf, 2010, p. 708. Italics added.).

Claims:

1. During the first stage of the demographic transition fertility and mortality rates were high.
2. Natural population growth (growth rate) was fluctuating around zero

In this quotation the "natural population growth" is identified as the *difference* between the fertility and mortality rates, i.e. as the *growth rate*. We shall recall that while the growth rate fluctuating around a constant value describes exponential growth, the growth rate "fluctuating around zero' describes the constant size of the growing entity, that is in our case, the constant size of the population. The claim made by Warf is contradicted by data, which show that for thousands of years the size of human population was *increasing*. The "natural population growth" (growth rate) could not have been "fluctuating around zero." Furthermore, two time-dependent quantities do not *have* to be large to make the difference between them small and fluctuating around zero, so the cause-effect relation is also incorrectly identified.

In line with the accepted interpretations of the first stage of the Demographic Transition, Lagerlöf writes:

> The Malthusian Regime in our model is a stable situation where death and birth rates are both high, and *population roughly constant*. Moreover, mortality is highly volatile, increasing dramatically in periods of big epidemic shocks. In periods with mild shocks population expands. This worsens the impact of the next epidemic, equilibrating population back to its Malthusian state (Lagerlöf, 2003a, p. 756. Italics added).

> In our model, the world can thus be stuck in a *Malthusian equilibrium* for centuries and then suddenly escape, and never contract back. As suggested by a referee, this process could possibly be interpreted in terms of wars, instead of epidemics (Lagerlöf, 2003a, p. 766. Italics added.).

*Throughout human history, epidemics, wars and famines have shaped the growth path of population.* Such shocks to mortality are the central theme of the model set up by Lagerlöf, which endogenously generates a long phase of *stagnant population* and living standards, followed by an industrial revolution and a demographic transition (Lagerlöf, 2003b, pp. 434, 435. Italics added.).

Claims:

1. It is assumed that there was a Malthusian regime
2. It is assumed that Malthusian regime is characterised by high birth and death rates
3. During the Malthusian regime population is roughly constant
4. Mortality is highly volatile
5. Mortality increases dramatically in periods of big epidemic shocks
6. Population expands when the mortality shocks are mild
7. Expanding population worsens the impact of the next epidemic and equilibrates population to the Malthusian state
8. Malthusian equilibrium lasts for centuries
9. The process of Malthusian equilibrium can be also explained by wars instead of epidemics
10. Throughout human history, epidemics, wars and famines have shaped the growth path of population.
11. Model based on the assumption of shocks to mortality generates a long phase of stagnant population

It would be interesting to see convincing support for all these claims. In order to study the impact of demographic crises on the growth of human population we have three options: (1) to carry out an extensive survey of recorded demographic crises and see whether they are in any way correlated with changes in the data representing the growth of human population; (2) to carry out an extensive survey of recorded demographic crises, incorporate them in a mathematical model and see whether the model generates significant changes in the calculated size of human population and whether the model-predicted distributions of the size of the population fit the empirical data; and (3) to incorporate random fluctuations in birth and death rates in a suitably designed model, without worrying whether they are in any way related to the recorded incidents of demographic catastrophes and see whether model-generated distributions for the size of the population show any fluctuations or clearly uneven variations and whether these distributions fit the relevant data.

Lagerlöf carried out a limited model-based study of the effects of random fluctuations by assuming them only for the death rates. Close examination of his results shows that they are *in contradiction* with his conclusions about the existence of the Epoch of Malthusian Stagnation and that they *fail to fit* the relevant population data.

It appears that Lagerlöf interpreted the *roughly constant growth* rate as the *roughly constant population*. When we look at the results of his Monte Carlo calculations (Lagerlöf, 2003b, p. 436) carried out using the model described in his companion publication (Lagerlöf, 2003a) we can see that *he displays roughly constant growth rates generated by his model but interprets them as population growth*.

As already mentioned earlier, roughly constant growth rates produce *exponential* growth. They do not produce roughly constant population. Thus, rather than showing that the population was roughly constant, fluctuating and stagnant, Lagerlöf has shown that his model generated *exponential growth of population*, which is definitely not stagnant. Paradoxically, therefore, Lagerlöf has shown that if we assume that the growth of population is controlled by random forces, we shall *not* produce a stagnant state with a roughly constant size of the population but a steady, non-stagnant, *exponential* growth. It is interesting that under similar conditions but using a different approach for their computer simulations, Artzrouni and Komlos (1985) also did not generate stagnation but a steady *exponential* growth of human population.

Lagerlöf as well as Artzrouni and Komlos were on the verge of making a breakthrough discovery. Had they carried out their research properly, had they adhered to the principles of impartial and unbiased scientific investigation, they would have discovered that the assumption of the existence of Malthusian oscillations resulted in producing a steadily increasing size of human population without any signs of oscillations or fluctuations. They would have discovered that their models strongly question the whole concept of Malthusian stagnation and that perhaps all these positive checks proposed by Malthus do not have such a profound effect on the growth of human population as feared by him and as accepted by so many people who do not seem to question his original suggestions and expectations.

However, results of Lagerlöf as well as of Artzrouni and Komlos also show that while producing exponential growth, as expected by Malthus, the generated distributions *do not fit the data*. This is another interesting and important clue. Perhaps Malthus was not correct in assuming that the population if unchecked increases exponentially. Perhaps he would have suggested something different if he had access to all the data so easily available to us. Perhaps we do not have to accept blindly and reverently all his concepts. Perhaps they are not immune to the process of scientific investigation; particularly that Malthus never expected or wanted to have such an unassailable immunity for his concepts.

It is hard to understand why Lagerlöf did not compare results of his model calculations with the data describing the growth of human population because he *had* access to the relevant data (Maddison, 2001). It was such an essential and important step to take but for whatever reason it was not taken.

> In our model, this leads to a *constant rate* of population growth prior to the adoption of the Solow technology. This result is consistent with population data from Michael Kremer (1993), where *the growth rate of population fluctuates around a small constant* throughout most of the Malthusian period (from 4000 B.C. to A.D. 1650)" Hansen & Prescott" (2002, p. 1205. Italics added.).

Claims:

1. Growth rate of population was constant during the Malthusian regime (i.e. prior to the adoption of Solow technology)
2. Constant growth rate is consistent with population data from Michael Kremer (1933)
3. Kremer's data show that the growth rate of population fluctuates around a small constant throughout most of the Malthusian period

*First*, as mentioned earlier, constant growth rate should not be confused with constant population. Constant growth rate produces exponential growth.

*Second*, we would have to show convincingly that the growth rate was indeed fluctuating around a small constant value. Kremer (1963) did not carry out an extensive study of the growth rate but his limited investigation shows that it was *not* constant and that it was *not* fluctuating, but that it was *increasing approximately linearly* with the *size* of the population. Hansen and Prescott must have seen these results because Kremer presents them in a graph, which is impossible to miss, and yet for some unexplained reason they did not use them.

The regularity noticed by Kremer is in perfect agreement with the evidence of the hyperbolic growth (Shklovskii, 1962, 2002; von Foerster & Amiot, 1960; von Hoerner, 1975). This combined evidence deserves further investigation but it strongly suggests that the growth rate was probably never fluctuating around a constant value.

It might not be immediately obvious but this short declaration that "the growth rate of population fluctuates around a small constant throughout most of the Malthusian period" contains a huge amount of questionable information. We would have to do a lot of hard work to be able to say so much, so categorically and with such a confidence.

We would have to prove convincingly that the Malthusian period existed. Such a proof, on its own, would have been a monumental achievement deserving a special recognition. We would then have to study the behaviour of the growth rate during that period, preferably going back to the dawn of our existence some 200,000 or 300,000 years ago and show convincingly that sometimes the growth rate was not fluctuating around a constant value but *most* of the time it did.

Such leaps of faith, such confident but strongly questionable declarations, such claims containing so much unproven assertions occur repeatedly in the descriptions of the concept of the Epoch of Malthusian Stagnation. When accepted by faith, they can be easily reinforced by other dubious concepts, one wrong step followed by another, leading to an increasing accumulation of incorrect ideas, to the development of a system based on misconceptions, to the narratives, which might be interesting and fascinating but leading away from discovering correct interpretations.

> If population *density* increases the mortality rate rises, equilibrating population back to the Malthusian trap (Lagerlöf, 2003a, p. 765. Italics added.).

Here we have an example of an interesting *detail* in the concept of the Epoch of Malthusian Stagnation, the detail containing huge amount of information. This statement introduces the concept of the dependence of mortality rates on the *density* of human population. It offers an *explanation* of the *mechanism* of the Malthusian trap, whose existence is not supported by the population data. It describes some kind of a general rule stating that the Malthusian trap is activated when the population *density*, not its size, reaches a certain limiting value.

There is no research confirming the described mechanism; no research showing how the growth of human population depends on its *density*. Even if we could show some isolated examples of density-dependent growth we would have to demonstrate that they apply to regional and global populations. The best data available to us show the *time*-dependence of the size of human population and there is nothing in them to suggest any form of *density*-dependence, let alone the existence of the Malthusian trap.

This statement is yet another example of the leaps of faith, of confident declarations requiring a huge amount of work to be accepted as a reliable contribution to science. The descriptions of the Epoch of Malthusian Stagnation are full of such unscientific declarations. Indeed, they are made of them.

Other terms used to describe the alleged stagnant and fluctuating state of growth during this mythical Epoch of Malthusian Stagnation are "equilibrium trap" or "population trap" (Leibenstein, 1957; Nelson, 1956), "multiple equilibria" or "poverty trap" (Wang, 2005).

The belief in the stagnant and fluctuating growth is so strong that mathematical models are deemed successful if they can generate the desired oscillations during this mythical Epoch of Malthusian Stagnation, and *no-one seems to care about taking the next and the most essential step to compare model calculations with the population data* (Galor, 2005; Galor & Weil, 2000; Lagerlöf, 2003a, 2003b). As long as oscillations of some kind are generated by a mathematical model, they are taken as the proof of the existence of the Epoch of Malthusian Stagnation. This line of reasoning shows that the primary, if not the exclusive, aim of such mathematical exercises is to translate a story into a mathematical language and when the translation is done correctly, when mathematical formulae generate *any kind of oscillations*, large or small, significant or negligible, these formulae are taken as a proof that the myth represents reality.

The Epoch of Malthusian Stagnation is also described as the Age of Pestilence and Famine (Omran 1971, 1983, 1986, 1998).

> In this stage, the major determinants of death are the Malthusian positive checks, namely epidemics, famines and wars (Omran, 1983, p. 306; Omran, 2005, p. 737).

> Even if fertility approached its biologic maximum, depopulation could and did occur as a result of epidemics, wars and famines, which repeatedly pushed mortality levels to high peaks (Omran, 2005, p. 733).

Claims:

1. During the Age of Pestilence and Famine (Epoch of Malthusian Stagnation) major determinants of death are the Malthusian positive checks (epidemics, famines and wars)
2. Depopulation was occurring even when fertility was approaching its biological maximum because epidemics, wars and famines were repeatedly pushing mortality levels to high peaks

To justify the first claim we would have to have reliable records of the *causes of death* over thousands of years. We would then have to show convincingly that indeed the major causes of death were epidemics, famines and wars. We would also have to show that there was a clear change in the causes of death when the Epoch of Malthusian Stagnation ceased to exist. We cannot present such proofs because we do not have the supporting data.

To justify the second claim we would have to have reliable records of fertility and mortality over thousands of years. We would then have to demonstrate that fertility was approaching biological limits, that such events were coinciding with high mortality peaks and that these high mortality peaks were caused by epidemics, wars and famines.

> During the first stage, *mortality vacillated at high levels*, with infectious disease as the main cause of death plus a large proportion due to wars and famines (Robine, 2001, p. 191. Italics added.).
>
> Claims:
>
> 1. During the first stage of demographic transitions mortality vacillated at high levels
> 2. The main causes of death were infectious diseases
> 3. Large proportion of death were caused by wars and famines

We cannot prove that "mortality vacillated at high levels" because we have no relevant data for "the first stage" to carry out such a study, the stage that is assumed to have lasted for thousands of years. We cannot prove that these imagined and strongly-desired vacillations were correlated with infectious disease, wars and famines. We cannot prove that the *main* causes of deaths were infectious diseases. We cannot prove that a *large propo*rtion of death was due to wars and famines. We do not have records of causes of death extending over thousands of years. We do not know how the causes of death were changing over time. We do not have the records to help us to distinguish between the major cause and secondary causes. We do not know whether the main cause of death was the same over thousands of years. In order to accept the concept of the Epoch of Malthusian Stagntion and all these claims have to be accepted by faith.

> The first transition phase, called the 'Age of Pestilence and Famine,' is characterized by *high and fluctuating mortality rates*, variable life expectancy with low average life span, and *periods of population growth that are not sustained* (McKeown, 2009, p. 20S. Italics added.).

Claims:

1. During the Age of Pestilence and Famine (Epoch of Malthusian Stagnation) mortality rates were high and fluctuating
2. Average life span was low
3. There were periods when the population growth was not sustained

Mortality rates might have been high and fluctuating but we have no data extending over thousands of years to prove it. Furthermore, we would yet have to show that these hypothetical high and fluctuating mortality rates could have been responsible for creating stagnation. The same applies to the low average life span. As for the "periods of population growth that are not sustained" we can easily demonstrate using the population data that this claim is not sustained.

> The positive forces of growth had existed all along. However, they had been counterbalanced by the negative forces of malnutrition and disease (Komlos & Baten, 2003, p. 19).

We have no reliable empirical evidence to support this claim, no study of positive and negative forces, no study of their balancing, and no study of their influence on the growth of human population. There is also no attempt to consult the population data (Maddison, 2010;

Manning, 2008; US Census Bureau, 2013) to test the concept of the balancing of positive and negative forces. Here again, and quite typically, an attractive declaration fitting the generally accepted concept is made without trying to support it by solid scientific evidence.

**Summary and conclusions**

We have given a few examples of unsubstantiated claims associated with the concept of the Epoch of Malthusian Stagnation, examples of confident declarations, which have to be accepted by faith. They represent only a part of a wider range of misconceptions about the growth of human population and about the related issue of the economic progress as expressed in terms of the Gross Domestic Product (GDP) or the GDP per capita.

It is impossible to correct the mistakes of the past 200 years in just one short article but we have mentioned a few ways of testing and correcting the prevailing misconceptions. In the next few articles we shall focus on some of these ways.

A huge step forward can be made if we identify and abandon incorrect concepts no matter how popular and how attractive they might appear to be. However, each impartial and unbiased examination of empirical evidence can also take us a step closer to a better and correct understanding of the human population dynamics and of the economic progress.

Guest, A. M., & Almgren, G. (2001). Demographic Transition. In *Encyclopedia of Sociology* (2nd ed.), Vol. 1, (pp. 621-630). Detroit: Macmillan Reference.

Habakkuk, H. J. (1953). English population in the eighteenth century. *Economic History Review*, *6*(2), 117-133

Hansen, G. D., & Prescott, E. C. (2002). Malthus to Solow. *The American Economic Review*, *92*(2), 1205-1217.

Haupt, A., & Kane, T. T. (2005). *Population handbook* (5th edn.). Washington, DC: Population Reference Bureau.

Johnson, C. N., & Brook, B. W. (2011). Reconstructing the dynamics of ancient human populations from radiocarbon dates: 10 000 years of population growth in Australia. *Proceedings of the Royal Society B, 278*, 3748-3754.

Kirk, D. (1996). Demographic Transition Theory. *Population Studies*, *50*, 361-387.

Komlos, J. H. (1989). Thinking about Industrial Revolution. *Journal of European Economic History, 18*, 191-206.

Komlos, J. H. (2000). The Industrial Revolution as the escape from the Malthusian Trap. *Journal of European Economic History, 29*, 307-331.

Komlos, J., & Baten, J. (2003). The Industrial Revolution as the escape from the Malthusian trap. *Munich Discussion Paper No. 2003-13*, Volkswirtschaftliche Fakultät, Ludwig-Maximilians-Universität München

Kremer, M. (1993). Population growth and technological change: One million B.C. to 1990. *Quarterly Journal of Economics*, *108*(3), 681–716.
38

43